\documentclass[twocolumn]{aastex7}

\usepackage{amsmath}

\begin{document}

\title{Partial Differentiation of Callisto as Possible Evidence for Pebble Accretion}

\author[orcid=0000-0003-2993-5312,sname='Shibaike']{Yuhito Shibaike}
\affiliation{Graduate School of Science and Engineering, Kagoshima University}
\email[show]{yuhito.shibaike@sci.kagoshima-u.ac.jp}


\begin{abstract}
``Planetesimal or pebble'' is one of the most fundamental open questions in planet formation theory. Similarly, ``satellitesimal or pebble'' remains unsettled regarding the formation of the Galilean satellites. I focus on a unique characteristic of Callisto—its interior is estimated to be only partially differentiated based on gravitational field measurements. I robustly demonstrate that such a state is not achievable through satellitesimal accretion, which inevitably leads to significant differentiation, but can be maintained with pebble accretion. Pebbles can release their impact energy at the surface of the satellite, allowing efficient radiative cooling, and their impact velocities can be reduced by aerodynamic drag from the circumplanetary gas disk. If future missions such as JUpiter ICy moons Explorer (JUICE) confirm that Callisto is indeed only partially differentiated, it could provide the first observed evidence for the pebble accretion mechanism—not only in the context of satellite formation, but also in the broader framework of planet formation.
\end{abstract}



\section{Introduction} \label{sec:Introduction}
Pebble accretion is an accretion mechanism of planets proposed as an alternative to the planetesimal accretion scenario in the context of planet formation theory \citep{orm10,lam12}. Dust particles in protoplanetary disks grow large toward planetesimals, km-sized building blocks of planets, by the repetition of their mutual collisions. However, the particles in the gas disks rotating with sub-Kepler due to the gas pressure have to receive the head wind from the gas, resulting in loss of their angular momentum and their spiral fall toward the central stars before they grow to planetesimals \citep{ada76,wei77}. Therefore, the alternative pebble accretion scenario has been proposed to avoid the issue; a small number of large bodies directly accrete the radially drifting particles called {\it pebbles}. This scenario opened a new window to form planets, and a lot of research has worked on the scenario so far. However, the question which accretion process dominates the planet formation is still unknown, because it depends on the size of pebbles and their size distribution, the amount of the solids and the disk model \citep[e.g.,][]{dra22,kes23,lyr23,orm24}. In this study, I propose a novel approach that can reveal the dominance of pebble accretion by focusing not on planetary formation but rather on the formation of a Jovian large satellite, Callisto.

The formation scenarios for large satellites rotating around gas giants have also two groups: satellitesimal accretion and pebble accretion. Satellitesimals are km-sized building blocks of satellites, the counterpart of planetesimals in the planet formation theory. The classical idea of the satellite formation is that a lot of satellites formed by accreting satellitesimals in circumplanetary gas disks, byproducts of the gas accretion of the parental gas giants, but fell onto the planets by Type I migration repeatedly, and the current satellites are the last generation of the repeatedly formed satellites surviving the disappearance of the gas disks \citep[e.g.,][]{can06,sas10}. However, this scenario has the same issue as the planetesimal accretion: dust particles in circumplanetary disks drift toward the planets as pebbles before they grow large to satellitesimals \citep{shi17}. Therefore, the pebble accretion scenarios for the satellite formation have also been proposed \citep{ron17,shi19,ron20}, but there has been no definitive answer for the question of the dominant mechanism of the formation like the situation of the planet formation.

Callisto is one of the four large icy satellites of Jupiter called Galilean satellites. Gravitational field measurements by the Galileo mission estimates Callisto's normalized moment of interior (MoI) as $C/(M_{\rm s}R_{\rm s}^{2})=0.3549\pm0.0042$ under the assumption of hydrostatic equilibrium, where $C$ is MoI, and $M_{\rm s}$ and $R_{\rm s}$ are the current mass and radius of the satellite \citep{and01}. This high MoI value suggests that Callisto's interior is only partially differentiated. The simplest interpretation is that Callisto has an outer icy shell with $\sim300~{\rm km}$ thickness ($\sim1/3$ of the whole satellite in mass) and a homogeneous ice and rock mixture under the shell \citep{sch04}. JUpiter ICy moons Explorer (JUICE) mission (launched in 2023 and starts Callisto flybys in 2031) will test the hydrostatic equilibrium hypothesis rather than assuming it and will provide more robust and precise estimates of the differentiation state of Callisto \citep{cap22}.

The conditions necessary to maintain Callisto's partially differentiated interior have been investigated within the framework of the satellitesimal accretion scenario \citep[e.g.,][]{bar08}, and they are found to be stringent. Satellitesimals deposit their impact energy under the surface of the satellites avoiding efficient radiative cooling from the surface. As a result, global interior melting must occur in large satellites like Callisto over a broad range of parameters \citep{mon14}. As larger impactors deliver more energy to the interior, increasing the proportion of large impactors during accretion leads to higher internal temperature and a greater extent of interior differentiation \citep{ben25}. On the other hand, the interior of Callisto can remain undifferentiated with the pebble accretion scenario when the accretion takes $\sim10~{\rm Myr}$, but only limited situations have been investigated so far \citep{shi19}.

In this work, I will investigate the heating of Callisto's interior with both the satellitesimal and pebble accretion scenarios with broad ranges of parameters and robustly show that the fundamental characteristics of pebbles maintain the partially differentiated interior; pebbles are too small to deposit their impact energy under the surface, and the gas around the satellite reduces the impact velocity of pebbles, resulting in less effective accretion heating.

\begin{table*}[htbp]
\centering
\caption{Properties of Callisto and Ganymede}
\label{tab:Callisto}
\hspace{-4.5em}  
\begin{tabular}{lccc}
\hline
Descriptions & Symbols & Callisto & Ganymede \\
\hline
Observed \\
\hline
Mass & $M_{\rm s}$ & $1.076\times10^{23}~{\rm kg}$ & $1.482\times10^{23}~{\rm kg}$ \\
Radius & $R_{\rm s}$ & $2410~{\rm km}$ & $2631~{\rm km}$ \\
Mean density & ${\bar\rho}$ & $1.835~{\rm g~cm}^{-3}$ & $1.943~{\rm g~cm}^{-3}$ \\
Orbital radius & $r$ & $26.4~R_{\rm J}$ & $15.0~R_{\rm J}$ \\
Normalized MoI & $C/(M_{\rm s}R_{\rm s}^{2})$ & $0.3549\pm0.0042$ & $0.3115\pm0.0028$  \\
\hline
Assumed \\
\hline
Specific heat & $C_{\rm p}$ & $1.48\times10^{7}~{\rm erg~g^{-1}~K^{-1}}$ & $1.37\times10^{7}~{\rm erg~g^{-1}~K^{-1}}$ \\
Rock mass fraction & $m_{\rm r}$ & $0.44$ & $0.52$ \\
Disk temperature & $T_{\rm disk}$ & $110~{\rm K}$ & $160~{\rm K}$ \\
Gas surface density & $\Sigma_{\rm g}$ & $9\times10^{3}~{\rm g~cm}^{-2}$ & $1.5\times10^{4}~{\rm g~cm}^{-2}$ \\
\hline
\end{tabular}
\end{table*}

\section{Methods} \label{sec:methods}
\subsection{Surface and subsurface temperature determined by accretion heating} \label{sec:temperature}
\begin{figure}[t]
\plotone{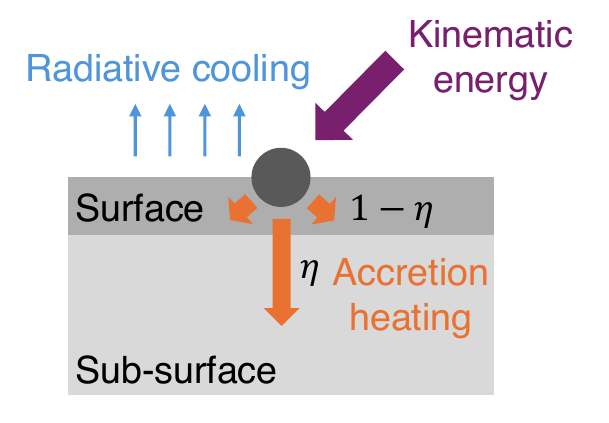}
\caption{Schematic picture of the accretion heating model. 
\label{fig:model}}
\end{figure}

Accretion of materials heats the surface of the satellite, while radiative cooling reduces the surface temperature. At the same time, some part of the kinematic energy of the impacts is deposited under the surface, where the radiative cooling does not work (Fig. \ref{fig:model}). When the fraction of the impact energy deposited under the surface of the satellite is defined as $\eta$, the energy balance at the surface is expressed as
\begin{equation}
{\bar\rho} C_{\rm p}(T_{\rm surf}-T_{\rm disk})\dfrac{dR}{dt}=\dfrac{1-\eta}{2}\dfrac{\dot{M}v_{\rm imp}^{2}}{4\pi R^{2}}-\sigma_{\rm SB}(T_{\rm surf}^{4}-T_{\rm disk}^{4}),
\label{Tsurf}
\end{equation}
where $T_{\rm surf}(R)$ is the surface temperature of the satellite, $R$ is the radius of the satellite at the time after CAI's formation $t$, $v_{\rm imp}$ is the impact velocity, and $\sigma_{\rm SB}$ is the Stefan-Boltzmann constant. The accretion rate of the satellite is $\dot{M}\equiv dM/dt=4\pi{\bar\rho}R^{2}dR/dt$, where $M$ is the satellite mass at time $t$. I use Callisto's values for the mean density and capacity of the satellite, ${\bar\rho}=1.834~{\rm g~cm}^{-3}$ and $C_{\rm p}=1.48\times10^{7}~{\rm erg~g^{-1}~K^{-1}}$. I assume the disk temperature around the satellite as $T_{\rm disk}=110~{\rm K}$ as the value for the location of Callisto \citep{shi19}. Since radiative cooling does not work under the surface,
\begin{equation}
{\bar\rho} C_{\rm p}(T_{\rm sub}-T_{\rm surf})\dfrac{dR}{dt}=\dfrac{\eta}{2}\dfrac{\dot{M}v_{\rm imp}^{2}}{4\pi R^{2}},
\label{Tsub}
\end{equation}
where $T_{\rm sub}(R)$ is the subsurface temperature of the satellite.

\subsection{Accretion of satellitesimals and pebbles} \label{sec:accretion}
Accretion heating depends on how materials of satellites (i.e., impactors) accrete onto the satellite. I consider two types of accretion mechanisms, satellitesimal and pebble accretion, and assume three sets of properties for each: for satellitesimal accretion, \texttt{Hot accretion}, \texttt{Cold accretion}, and \texttt{Cold-limit accretion}, and for pebble accretion, \texttt{Large pebbles}, \texttt{Small pebbles}, and \texttt{Fragmentation-limited}.

The first property is $\dot{M}$, which is a function of $M$. When it is expressed as $\dot{M}\propto M^{q}$ for the whole accretion process with an index parameter $q$ ($q\neq1$), the accretion rate is 
\begin{equation}
\dot{M}=\dfrac{M_{\rm s}^{1-q}-M_{0}^{1-q}}{\tau_{\rm form}(1-q)}M^{q},
\label{Mdot}
\end{equation}
where $M_{0}$ is the initial mass of the satellite. I assume $M_{0}=3\times10^{23}~{\rm g}$, where effective pebble accretion starts roughly from that mass \citep{shi19}. I also define a parameter, formation period $\tau_{\rm form}\equiv t_{\rm end}-t_{\rm start}$, as the period between the start of formation (i.e., accretion) $t_{\rm start}$ and the end of formation $t_{\rm end}$. The satellite mass and the accretion timescale are then
\begin{equation}
M=\left\{\dfrac{t-t_{\rm start}}{\tau_{\rm form}}(M_{\rm s}^{1-q}-M_{0}^{1-q})+M_{0}^{1-q}\right\}^{1/(1-q)},
\label{M}
\end{equation}
and
\begin{equation}
\tau_{\rm acc}\equiv\dfrac{M}{\dot{M}}=\dfrac{(1-q)M^{1-q}}{M_{\rm s}^{1-q}-M_{0}^{1-q}}\tau_{\rm form}.
\label{tau-acc}
\end{equation}

In the case of satellitesimal accretion, the accretion mode of a satellite shifts from ``runaway growth'' to ``oligarchic growth'' as the satellite increases in mass, and most of its mass is accreted as the latter \citep{kok98,sas10}. I here only consider the oligarchic growth, so that the accretion rate for satellitesimal accretion is $\dot{M}\propto M^{2/3}$ \citep{kok98}. This is different from the assumptions of the previous works, $\dot{M}\equiv M/\tau_{\rm acc,const}\propto M$ \citep{bar08,ben25}. In the case of pebble accretion, the accretion also shifts from ``three-dimensional (3D)'' to ``two-dimensional (2D)'' accretion modes as the satellite grows massive \citep{ida16a}. This is because the accretion radius becomes larger than the pebble scale height at some point. Most of the mass of the satellite is accreted in the 2D mode \citep{shi19}. I here only consider the 2D accretion mode, so that the accretion rate for pebble accretion is $\dot{M}\propto M^{2/3}$ \citep{ida16a}. As a result, the index is $q=2/3$ in the both accretion mechanisms.

The second property is $\eta$, which depends on the accretion mechanism. In the case of satellitesimal accretion, a part of the impact energy is deposited under the surface of the satellite, depending on the size of the impactors. The exact value of $\eta$ is still unknown, but \citet{mon14} argues that the value takes from $0.1$ to $0.3$. Here, I set two cases, \texttt{Hot accretion} with $\eta=0.3$ and \texttt{Cold accretion} with $0.1$, based on the argument. In \citet{bar08}, the subsurface depositing effect is ignored, so that I also set \texttt{Cold-limit accretion} with $\eta=0$ as an extreme case. In the case of pebble accretion, I also assume $\eta=0$, because the subsurface energy deposition must be negligible. The size of pebbles in circumplanetary disks is $\lesssim1~{\rm m}$ \citep{shi17}, and the impact energy of such small impactors is deposited at the surface \citep{ben25}.

The third property is $v_{\rm imp}$, which also depends on the accretion mechanism. In the case of satellitesimal accretion, the velocity is
\begin{equation}
v_{\rm imp}=\sqrt{v_{\rm esc}^{2}+v_{\rm rel}^{2}},
\label{vimp-sat}
\end{equation}
where $v_{\rm esc}$ and $v_{\rm rel}$ are the escape and characteristic relative velocities, respectively. The escape velocity is
\begin{equation}
v_{\rm esc}=\sqrt{\dfrac{2GM}{R}},
\label{vesc}
\end{equation}
where $G$ is the gravitational constant. In the case of oligarchic growth, effects of viscous stirring and gas drag are balanced, the characteristic relative velocity is \citep{kok02}
\begin{equation}
v_{\rm rel}=\left(\dfrac{9M_{\rm sts}\ln\Lambda}{5\pi C_{\rm D}R_{\rm sts}^{2}r\rho_{\rm g}}\right)^{1/5}R_{\rm H}\Omega_{\rm K},
\label{vrel-sat}
\end{equation}
where $M_{\rm sts}$ and $R_{\rm sts}$ are the mass and radii of the satellitesimals. The Hill radius of the satellite and the Keplerian velocity are $R_{\rm H}=(M/(3M_{\rm p}))^{1/3}r$ and $\Omega_{\rm K}=\sqrt{GM_{\rm p}/r^{3}}$, where $M_{\rm p}=1~M_{\rm J}$ (Jupiter mass) is the planet mass, and $r=26.4~R_{\rm J}$ (Jupiter radius) is the orbital radius of the satellite (Callisto). The midplane gas density of the circumplanetary disk is $\rho_{\rm g}=\Sigma_{\rm g}/(\sqrt{2\pi}H_{\rm g})$, where $H_{\rm g}=c_{\rm s}/\Omega_{\rm K}$ is the gas scale height. The sound velocity is $c_{\rm s}=\sqrt{k_{\rm B}T_{\rm disk}/m_{\rm g}}$, where $k_{\rm B}$ is the Boltzmann constant, and $m_{\rm g}=3.9\times10^{-24}{\rm g}$ is the mean molecular mass of the gas. The Coulomb logarithm is $\ln\Lambda=3$ \citep{ste00}, and the drag coefficient is $C_{\rm D}=0.47$ \citep{per11}. The mass of satellitesimals is $M_{\rm sts}=4\pi/3\rho_{\rm sts}R_{\rm sts}^{3}$, where the radii and density of satellitesimals are assumed as $R_{\rm sts}=1~{\rm km}$ and $\rho_{\rm sts}=\rho_{\rm s}$. I also assume the gas surface density of the circumplanetary disk at the location of Callisto as $\Sigma_{\rm g}=9\times10^{3}~{\rm g~cm}^{-2}$ \citep{shi19}. Note that these assumptions do not alter the conclusions of this paper (see Section \ref{sec:gasdrag}).

On the other hand, the impact velocity of pebbles is significantly affected by the gas drag of the circumplanetary disk depending on their Stokes numbers ${\rm St}$,
\begin{equation}
v_{\rm imp}=\min\left(\sqrt{v_{\rm esc}^{2}+v_{\rm rel}^{2}},v_{\rm set}\right),
\label{vimp-peb}
\end{equation}
where the settling velocity is
\begin{equation}
v_{\rm set}=gt_{\rm s}.
\label{vset}
\end{equation}
The gravitational acceleration at the surface of the satellite is $g=GM/R^{2}$, and the stopping time of the pebbles is $t_{\rm s}\equiv{\rm St}/\Omega_{\rm K}$, where ${\rm St}$ is the Stokes number of pebbles. The characteristic relative velocity of pebbles is
\begin{equation}
v_{\rm rel}=\left\{1+5.7\left(\dfrac{M/M_{\rm p}}{\eta_{\rm hw}^{3}/{\rm St}}\right)\right\}^{-1}v_{\rm hw}+v_{\rm sh},
\label{vrel-peb}
\end{equation}
where $v_{\rm hw}=\eta_{\rm hw}v_{\rm K}$ and $v_{\rm sh}=0.52(M/M_{\rm p}{\rm St})^{1/3}$ are the headwind and Keplerian shear velocities, respectively \citep{liu18,shi19}. The Keplerian velocity is $v_{\rm K}=r\Omega_{\rm K}$, and the headwind prefactor is
\begin{equation}
\eta_{\rm hw}=-\dfrac{1}{2}\left(\dfrac{H_{\rm g}}{r}\right)^{2}\dfrac{\partial\ln{\rho_{\rm g}c_{\rm s}^{2}}}{\partial\ln{r}}.
\label{eta-hw}
\end{equation}

Previous studies have shown that the Stokes number of pebbles is typically ${\rm St}\sim0.01-0.1$ when it is determined by radial drift \citep{shi17,shi19}. The Stokes number of pebbles is also determined by their fragmentation. The fragmentation-limited Stokes number ${\rm St_{frag}}$ is determined by the critical fragmentation velocity of their mutual collisions $v_{\rm cr}$, where their collision velocity is determined by the strength of turbulence $\alpha$ and the degree of radial drift (i.e., $\eta_{\rm hw}$). The fragmentation-limited Stokes number around Callisto is then \citep{oku16,shi19}
\begin{equation}
{\rm St_{frag}}=\dfrac{-3\alpha c_{\rm s}^{2}+\sqrt{9\alpha^{2}c_{\rm s}^{4}+4\eta_{\rm hw}^{2}v_{\rm K}^{2}v_{\rm cr}^{2}}}{2\eta_{\rm hw}^{2}v_{\rm K}^{2}}=0.0014
\label{St-frag}
\end{equation}
where I assume $\alpha=10^{-4}$ and $v_{\rm cr}=0.5~{\rm m~s}^{-1}$. This critical velocity is lower than the assumption by \citet{shi19}, but such a ``fragile'' threshold has recently been frequently inferred from observations of protoplanetary disks \citep{oku19,ued24}. Therefore, I consider the following three cases: \texttt{Large pebbles} with ${\rm St}=0.1$, \texttt{Small pebbles} with ${\rm St}=0.01$, and \texttt{Fragmentation-limited} with ${\rm St}=0.001$.

\subsection{Radiogenic heating} \label{sec:radiogenic}
The other dominant heating mechanism of satellites' interiors is radiogenic heating. There are various radioisotopes heating up the materials around them, but I only calculate the radiogenic heating by $^{26}$Al, which provides a factor of $\times100$ more heat than other short-lived radioisotopes \citep{bar08}. The temperature at the location $R_{\rm int}$ (i.e., distance to the center of the satellite) and time $t$ is then
\begin{equation}
T(R_{\rm int},t)=T_{\rm sub}(R_{\rm int})+\Delta T(R_{\rm int},t).
\label{T}
\end{equation}
The temperature increase by $^{26}$Al heating is
\begin{equation}
\begin{split}
\Delta T(R_{\rm int},t)&=\dfrac{1}{C_{\rm p}}\int_{t_{\rm f}(R_{\rm int})}^{t}m_{\rm r}q_{26}(t)dt \\
&=\dfrac{m_{\rm r}q_{26}(0)}{C_{\rm p}\lambda_{26}}\left[\exp(-\lambda_{26}t_{\rm f}(R_{\rm int}))-\exp(-\lambda_{26}t)\right],
\label{DeltaT}
\end{split}
\end{equation}
where $t_{\rm f}(R_{\rm int})$ is the formation time of the layer at $R_{\rm int}$ (i.e., $t_{\rm f}(R)=t$). The rock mass fraction of Callisto is assumed as $m_{\rm r}=0.44$ \citep{bar08}. The initial heating rate by $^{26}$Al is $q_{26}(0)=1.82\times10^{-7}~{\rm W~kg}^{-1}$ \citep{bar08} (see Appendix \ref{sec:half-26Al} for the detailed discussion). The heating rate decay is expressed as $\exp(-\lambda_{26}t)$ with $\lambda_{26}=9.68\times10^{-7}~{\rm year}^{-1}$ corresponding to a half life of $0.716~{\rm Myr}$. In every timestep, I first calculate $T_{\rm surf}$ with Eq. (\ref{Tsurf}) and then calculate $T_{\rm sub}$ with Eq. (\ref{Tsub}) and finally calculate $T$ with Eq. (\ref{DeltaT}). When $t\rightarrow\infty$, the current internal temperature is obtained,
\begin{equation}
\begin{split}
T_{\rm cur}(R_{\rm int})&=T_{\rm sub}(R_{\rm int})+\dfrac{1}{C_{\rm p}}\int_{t_{\rm f}(R_{\rm int})}^{\infty}m_{\rm r}q_{26}(t)dt \\
&=T_{\rm sub}(R_{\rm int})+\dfrac{m_{\rm r}q_{26}(0)}{C_{\rm p}\lambda_{26}}\exp(-\lambda_{26}t_{\rm f}(R_{\rm int})).
\label{Tcur}
\end{split}
\end{equation}
I calculate the melting temperature of ice in the satellite $T_{\rm melt}$ (see Appendix \ref{sec:melting}) and discuss the differentiation of the interior. I define the melt mass fraction of the satellite as
\begin{equation}
m_{\rm melt}\equiv\dfrac{1}{M_{\rm s}}\int^{R_{\rm s}}_{R_{0}}
f(R_{\rm int})dR_{\rm int},
\end{equation}
where
\begin{equation}
f(R_{\rm int})=
\begin{cases}
4\pi{\bar\rho}R_{\rm int}^{2} & \text{if}~^{\exists}t\in[t_{\rm start}, t_{\rm end}] \\
& \text{s.t.}~T(R_{\rm int},t)>T_{\rm melt}(R_{\rm int},t), \\
4\pi{\bar\rho}R_{\rm int}^{2} & \text{if}~T_{\rm cur}(R_{\rm int})>T_{\rm melt}(R_{\rm int},t_{\rm end}), \\
0 & \text{otherwise}.
\end{cases}
\end{equation}
The initial radius of the satellite is $R_{0}=(3M_{0}/(4\pi{\bar\rho}))^{1/3}$.

\section{Results} \label{sec:results}
\subsection{Internal temperature distribution of Callisto} \label{sec:temperature-Callisto}
First, I calculate the internal temperature of Callisto and compare it with the melting point of ice with fixed formation period $\tau_{\rm form}$ and starting time $t_{\rm start}$.

\begin{figure*}[htbp]
\centering
\includegraphics[width=0.95\linewidth]{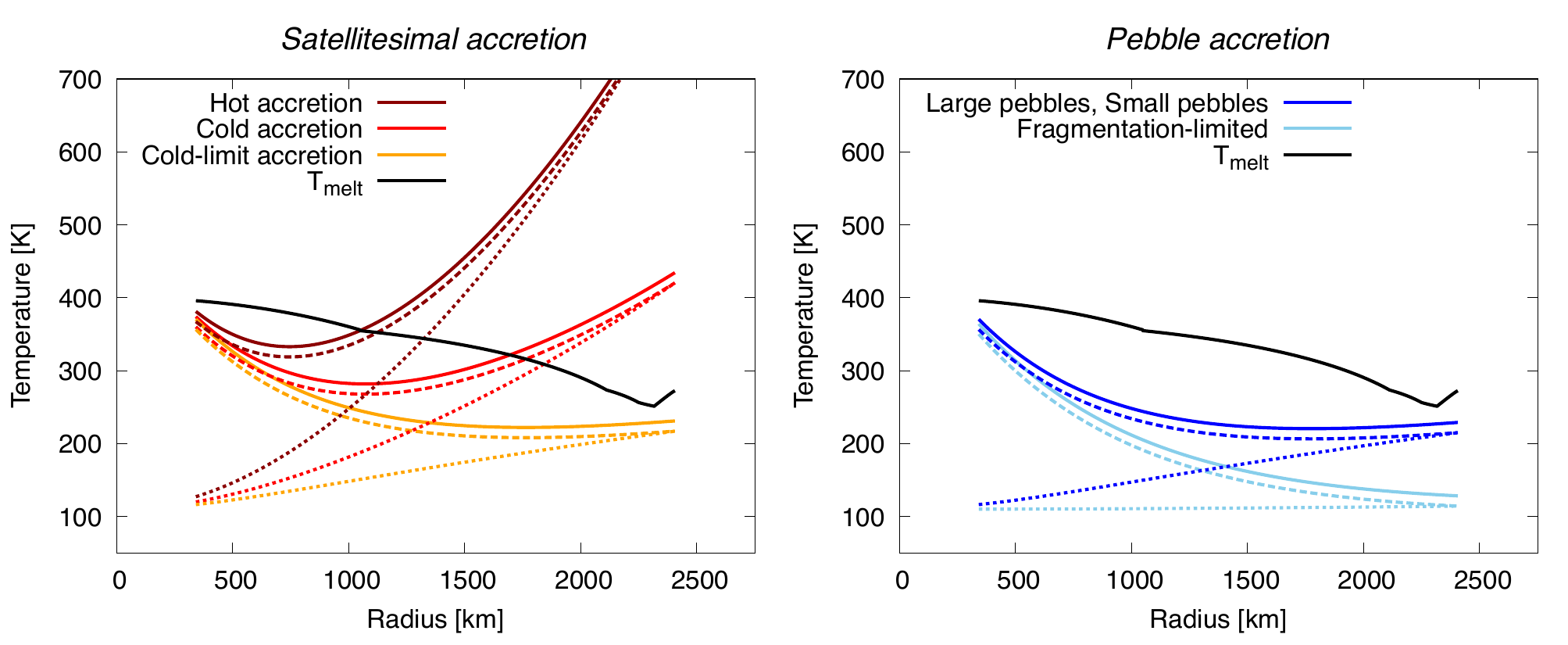}
\caption{Internal temperature distribution of Callisto with $\tau_{\rm form}=3~{\rm Myr}$, $t_{\rm start}=2~{\rm Myr}$, and $T_{\rm disk}=110~{\rm K}$. The left and right panels represent the temperature with satellitesimal and pebble accretion cases. In both panels, the solid, dashed, and dotted curves represent the current temperature $T_{\rm cur}$, temperature at the end of the formation $T(t_{\rm end})$, and temperature determined by accretion heating $T_{\rm sub}(t_{\rm end})$, respectively. The dark-red, red, orange curves in the left panel are the temperature with \texttt{Hot accretion} ($\eta=0.3$), \texttt{Cold accretion} ($\eta=0.1$), and \texttt{Cold-limit accretion} ($\eta=0$), respectively. The blue curves in the right panel represent the temperature with \texttt{Large pebbles} (${\rm St}=0.1$) and \texttt{Small pebbles} (${\rm St}=0.01$), where the temperature distribution of the two cases is overlapped completely. The sky-blue curves are the temperature with \texttt{Fragmentation-limited} (${\rm St}=0.001$). The black curves in the two panels are the melting points of ice at the end of the formation $T_{\rm melt}(t_{\rm end})$. Note that $T_{\rm melt}(t)$ is not a constant, because it depends on the pressure (see Appendix \ref{sec:melting}). 
\label{fig:temperature_Callisto_Callisto}}
\end{figure*}

The left panel of Fig. \ref{fig:temperature_Callisto_Callisto} shows that the temperature at the outer part of the satellite is high when $\eta$ is large. The dotted curves show that the temperature of that part is determined by accretion heating. As a result, the outer part of Callisto is melted with satellitesimal accretion in \texttt{Hot accretion} ($\eta=0.3$) and \texttt{Cold accretion} ($\eta=0.1$). Although the temperature can keep lower than the melting point in \texttt{Cold-limit accretion}, the assumption of this case, $\eta=0$, should not be realistic for satellitesimal impacts \citep{mon14}.

On the other hand, the temperature is lower than the melting point in any cases of pebble accretion. In \texttt{Large pebbles} and \texttt{Small pebbles}, the temperature distribution is the same as that of \texttt{Cold-limit accretion}, where all these cases assume $\eta=0$. In \texttt{Fragmentation-limited}, accretion heating does not increase the interior temperature, because the impact velocity is reduced significantly by the gas drag from the circumplanetary disk (see Section \ref{sec:gasdrag}). As a result, the temperature distribution is lower than that of \texttt{Large pebbles} and \texttt{Small pebbles} at any locations inside the satellite.

The increase of the temperature at the inner part of the satellite in any case is caused by radiogenic heating of $^{26}$Al. With $\tau_{\rm form}=3~{\rm Myr}$ and $t_{\rm start}=2~{\rm Myr}$, the radiogenic heating mainly occurs during the accretion process, which is understandable from the short half life of $^{26}$Al, $0.716~{\rm Myr}$. Also, the increased temperature does not reach the melting point of ice. 

\subsection{Melt mass fraction of Callisto} \label{sec:meltmassfraction}
I then vary $\tau_{\rm form}$ and $t_{\rm start}$ over the realistically broadest ranges; $0.5~{\rm Myr}\leq\tau_{\rm form}\leq20~{\rm Myr}$, where protoplanetary disks with one solar mass star could survive $\lesssim20~{\rm Myr}$ \citep{nak23}, and $0.5~{\rm Myr}\leq t_{\rm start}\leq10~{\rm Myr}$.

\begin{figure*}[htbp]
\centering
\includegraphics[width=0.9\linewidth]{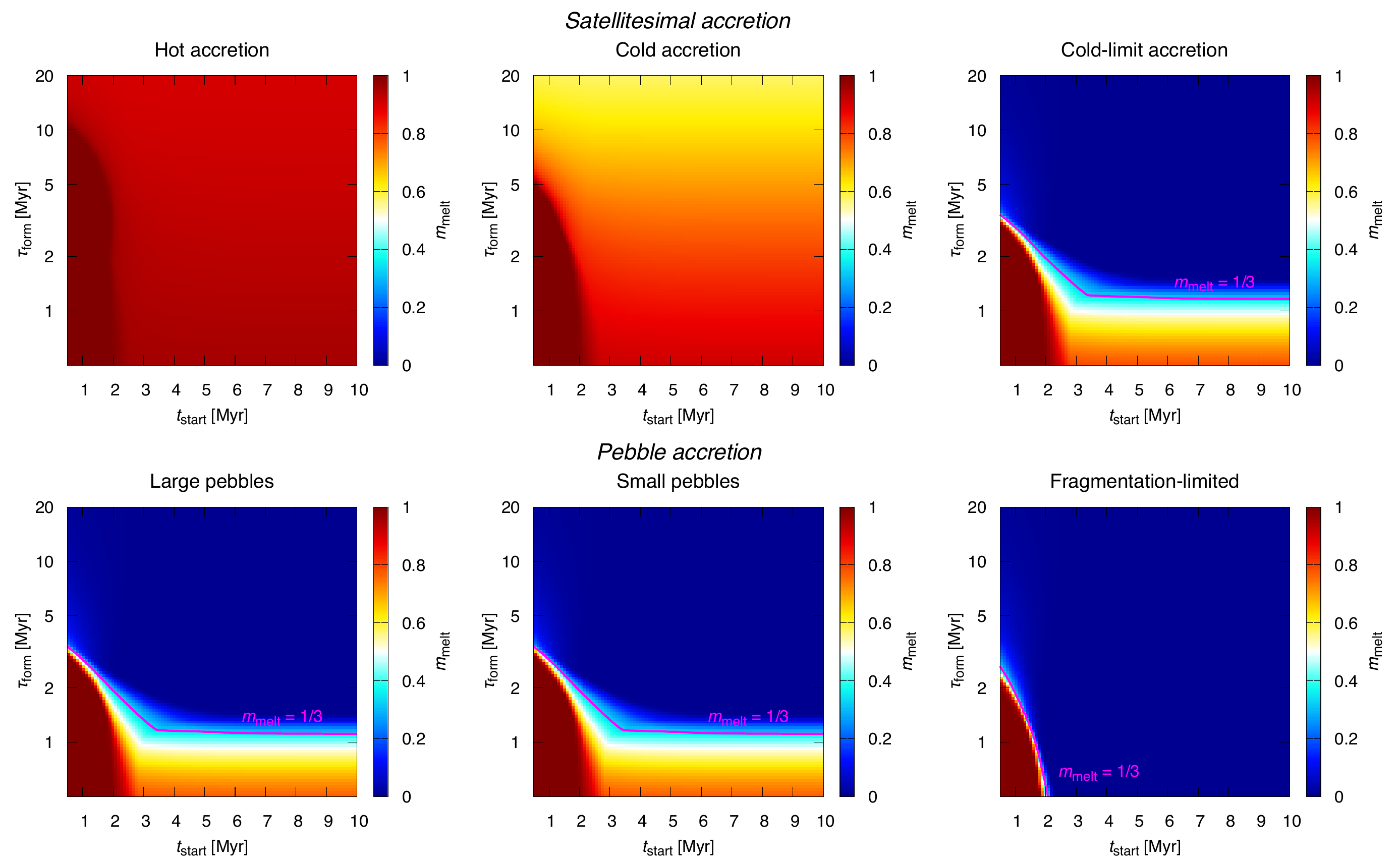}
\caption{Melt mass fractions of Callisto with the different accretion mechanisms. Satellitesimal accretion cases are \texttt{Hot accretion} ($\eta=0.3$), \texttt{Cold accretion} ($\eta=0.1$), and \texttt{Cold-limit accretion} ($\eta=0$). Pebble accretion cases are \texttt{Large pebbles} (${\rm St}=0.1$), \texttt{Small pebbles} (${\rm St}=0.01$), and \texttt{Fragmentation-limited} (${\rm St}=0.001$). The magenta curves represent $m_{\rm melt}=1/3$.
\label{fig:meltmassfraction_Callisto}}
\end{figure*}

Figure \ref{fig:meltmassfraction_Callisto} shows the melt mass fractions $m_{\rm melt}$ with the satellitesimal and pebble accretion cases. More than $90~{\rm wt}\%$ of Callisto is differentiated in \texttt{Hot accretion} and more than $50~{\rm wt}\%$ in \texttt{Cold accretion} for any set of $\tau_{\rm form}$ and $t_{\rm start}$. In those cases, the estimated $m_{\rm melt}$ of Callisto by the Galileo observations, $m_{\rm melt}\sim1/3$ \citep{sch04}, cannot be explained. The reason of the differentiation is accretion heating in most of the cases. The radiogenic heating by $^{26}$Al is effective only when $t_{\rm start}\lesssim2~{\rm Myr}$ and $\tau_{\rm form}\lesssim10~{\rm Myr}$ (\texttt{Hot accretion}) or $t_{\rm start}\lesssim2~{\rm Myr}$ and $\tau_{\rm form}\lesssim5~{\rm Myr}$ (\texttt{Cold accretion}), where they are represented as dark red regions in the panels. Only in the extreme case, \texttt{Cold-limit accretion}, the interior remains partially differentiated ($m_{\rm melt}\le1/3$) when $\tau_{\rm form}\geq1.2~{\rm Myr}$ except when the radiogenic heating is effective. I also find that Callisto is fully undifferentiated ($m_{\rm melt}=0$) if $\tau_{\rm form}\gtrsim1.5~{\rm Myr}$ and $t_{\rm start}\gtrsim2~{\rm Myr}$. The condition to maintain the partial differentiation, $\tau_{\rm form}\geq1.2~{\rm Myr}$, corresponds to $\tau_{\rm acc}\geq0.6~{\rm Myr}$ when $M=M_{\rm s}$ is substituted for Eq. (\ref{tau-acc}), which is consistent with the results of previous works \citep{bar08,ben25}.

On the other hand, in the cases of pebble accretion, the interior of Callisto robustly remains partially differentiated over broad ranges of parameters. Figure \ref{fig:meltmassfraction_Callisto} shows that the distribution of the melt mass fractions of \texttt{Large pebbles} and \texttt{Small pebbles} is almost the same as that of \texttt{Cold-limit accretion}. This is because the impact velocity is almost $v_{\rm imp}=v_{\rm esc}$, where $v_{\rm rel}$ is negligible in those cases (see Section \ref{sec:gasdrag}). In \texttt{Fragmentation-limited}, the interior remains partially differentiated with any $\tau_{\rm form}$ and $t_{\rm start}$ except for the effective radiogenic heating cases. This is because the accretion heating is not effective due to the aerodynamic drag (see Section \ref{sec:gasdrag}). I also find that Callisto is fully undifferentiated with any $\tau_{\rm form}$ if $t_{\rm start}\gtrsim2~{\rm Myr}$.

In summary, Fig. \ref{fig:meltmassfraction_Callisto} robustly shows that the formation of Callisto with a partially differentiated state can only be achieved by pebble accretion. To avoid significant differentiation with satellitesimal accretion, the extreme situation, $\eta=0$, must be assumed.

\section{Discussion} \label{sec:discussion}
\subsection{Effects of aerodynamic drag on pebbles} \label{sec:gasdrag}
\begin{figure}[t]
\centering
\includegraphics[width=0.95\linewidth]{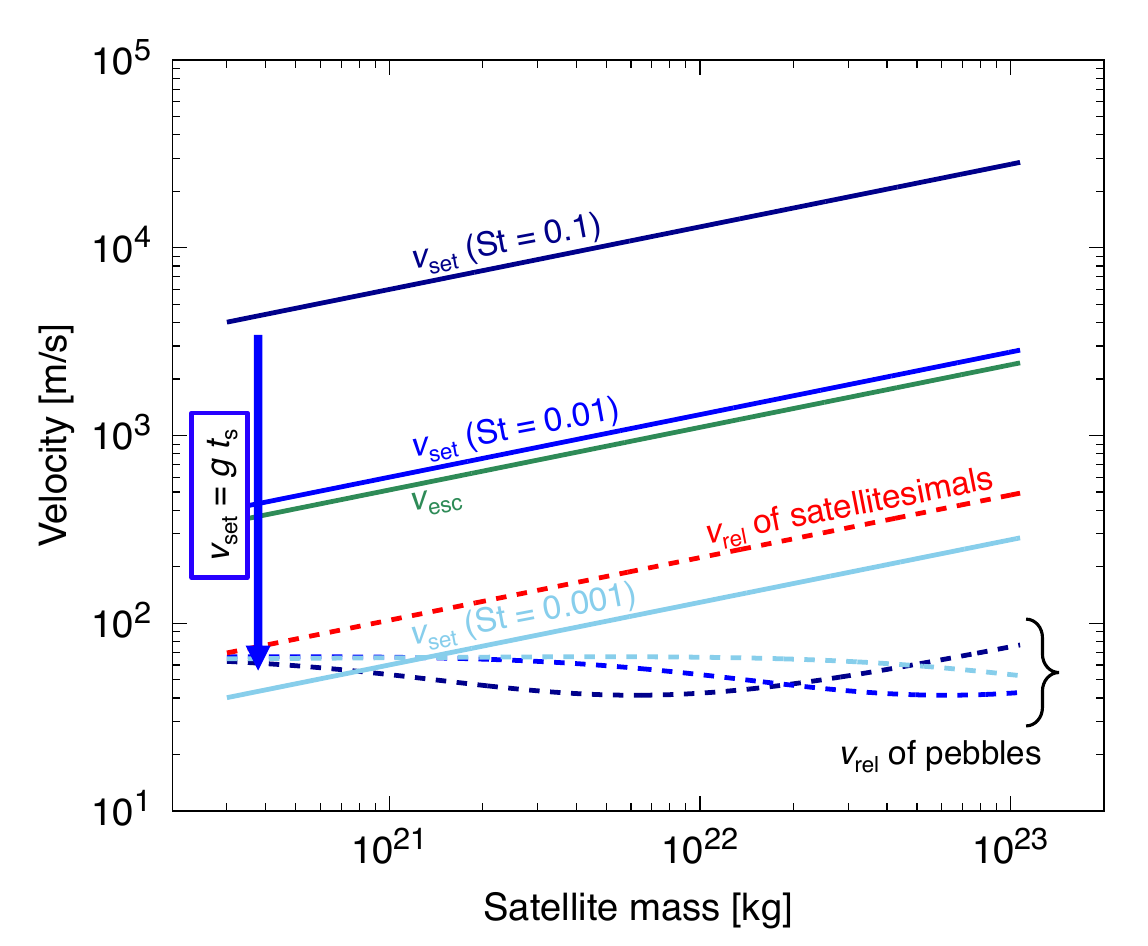}
\caption{Velocities of satellitesimals and pebbles accreting onto Callisto. The dark-blue, blue, and sky-blue solid lines represent the settling velocities of pebbles ($v_{\rm set}=gt_{\rm s}$) in \texttt{Large pebbles} (${\rm St}=0.1$), \texttt{Small pebbles} (${\rm St}=0.01$), and \texttt{Fragmentation-limited} (${\rm St}=0.001$), respectively. The blue dashed curves are those characteristic relative velocities $v_{\rm rel}$. The red dashed line is $v_{\rm rel}$ of satellitesimals. The green solid line is the escape velocity $v_{\rm esc}$. Note that the pebbles cannot fall onto Callisto faster than $\sqrt{v_{\rm esc}^{2}+v_{\rm rel}^{2}}$ (see Eq. (\ref{vimp-peb})), and thus $v_{\rm set}$ for ${\rm St}=0.1$ and $0.01$ is not the actual physical velocity they acquire.
\label{fig:mass-vel_Callisto}}
\end{figure}

Aerodynamic drag reduces the impact velocity $v_{\rm imp}$ of pebbles, resulting in the undifferentiated formation of Callisto in \texttt{Fragmentation-limited}. Figure \ref{fig:mass-vel_Callisto} shows that the settling velocity $v_{\rm set}$ of the pebbles is small as the Stokes number is small, which is clear from Eq. (\ref{vset}) (see the blue arrow). When ${\rm St}=0.1$ and $0.01$, $v_{\rm set}$ is faster than the escape velocity $v_{\rm esc}$, so that $v_{\rm imp}$ does not depend on $v_{\rm set}$ (Eq. (\ref{vimp-peb})). The figure also shows that $v_{\rm rel}$ is much slower than $v_{\rm esc}$ in any case. Therefore, the impact velocity is determined by $v_{\rm esc}$ (Eq.(\ref{vimp-peb})), resulting in the same $v_{\rm imp}$ as that of satellitesimal accretion. Therefore, the internal temperature distribution (Fig. \ref{fig:temperature_Callisto_Callisto}) and the melt mass fractions (Fig. \ref{fig:meltmassfraction_Callisto}) of \texttt{Large pebble} and \texttt{Small pebbles} are the same as those of \texttt{Cold-limit accretion}. However, when ${\rm St}=0.001$, $v_{\rm set}$ is much slower than $v_{\rm esc}$, resulting in significantly slower impact velocity. As a result, the internal temperature distribution is lower than the others (Fig. \ref{fig:temperature_Callisto_Callisto}), and $m_{\rm melt}$ is small except when the radiogenic heating is effective (Fig. \ref{fig:meltmassfraction_Callisto}).

The critical Stokes number that the aerodynamic drag is effective is derived from $v_{\rm esc}=v_{\rm set}$ (Eqs. (\ref{vesc}) and (\ref{vset})),
\begin{equation}
\begin{split}
{\rm St_{crit}}&=2\left(\dfrac{v_{\rm esc}}{v_{\rm K}}\right)^{-1}\left(\dfrac{R}{r}\right) \\
&=0.0086\left(\dfrac{M_{\rm p}}{1~M_{\rm J}}\right)^{1/2}\left(\dfrac{{\bar\rho}}{1.835~{\rm g~cm}^{-3}}\right)^{-1/2} \\
&~~~\times\left(\dfrac{r}{26.4~R_{\rm J}}\right)^{-3/2}. \\
\label{St-crit}
\end{split}
\end{equation}
Therefore, aerodynamic drag only plays a role in \texttt{Fragmentation-limited} in the case of Callisto. Also, the critical Stokes number only depends on the orbital radii of the satellites if they are in the same system and have the same density (i.e, composition).

\subsection{Comparison to Ganymede} \label{sec:Ganymede}
Ganymede is another large icy satellite of Jupiter with slightly larger mass and density than those of Callisto (Table \ref{tab:Callisto}). The normalized MoI is smaller, suggesting that the interior of Ganymede is fully differentiated. The reason why only Ganymede achieved the full differentiation has been investigated a lot \citep{bar08,bar10,shi19,ben25}, but there have been no clear answer yet.

\begin{figure*}[htbp]
\centering
\includegraphics[width=0.9\linewidth]{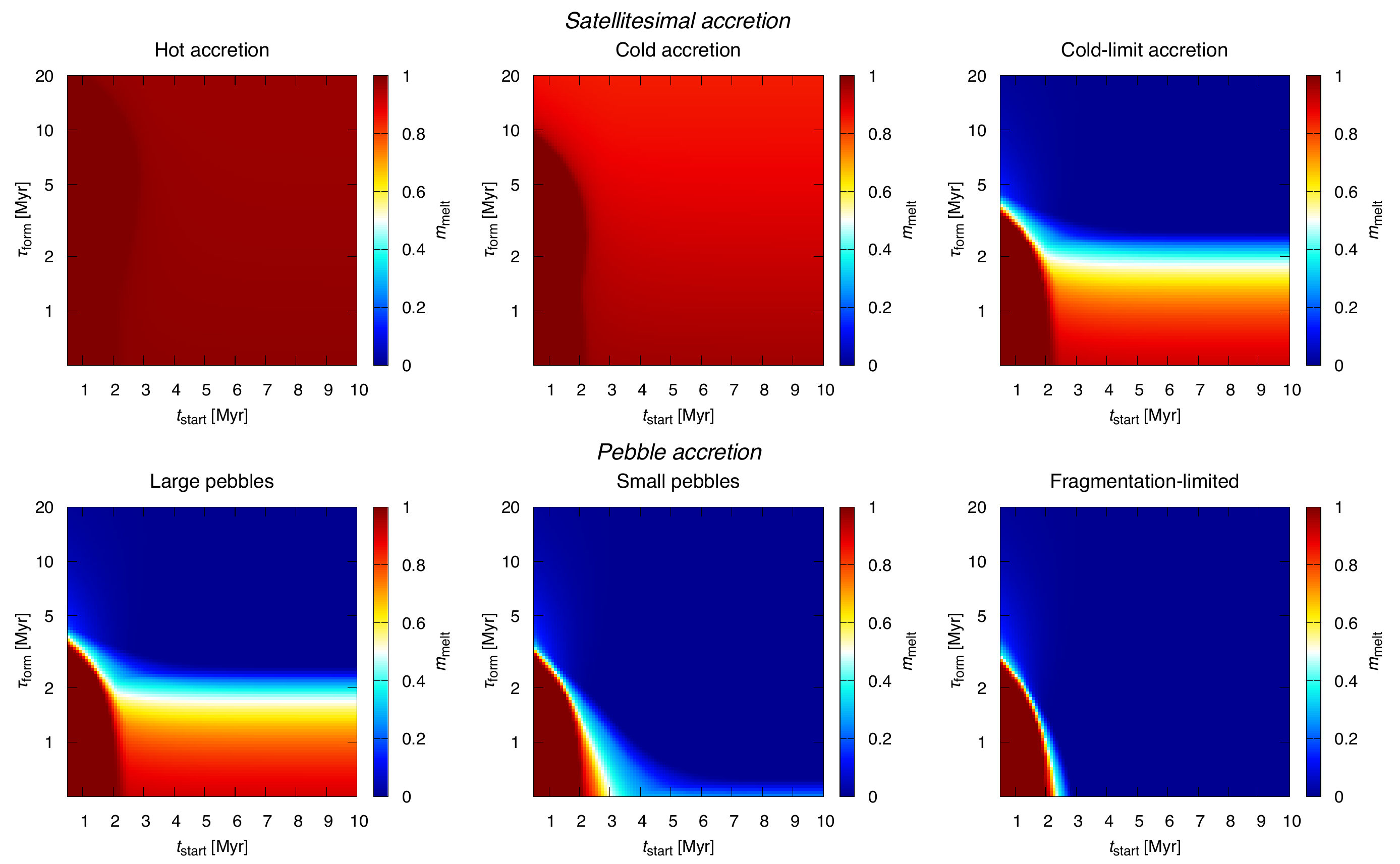}
\caption{Melt mass fractions of Ganymede. The properties of Ganymede are listed in Table \ref{tab:Callisto}.}
\label{fig:meltmassfraction_Ganymede}
\end{figure*}

Figure \ref{fig:meltmassfraction_Ganymede} shows that the trends of the melt mass fraction of Ganymede are the same as those of Callisto. In the case of pebble accretion, the melt mass fraction of Ganymede is $m_{\rm melt}\lesssim0.9$ except when the radiogenic heating by $^{26}$Al is effective. Therefore, it is necessary to melt the interior with subsequent processes in order to reproduce the full differentiation of Ganymede within the pebble accretion scenario. The long-term phenomena discussed in Section \ref{sec:ignore} may contribute to explain it, but its investigation is beyond the scope of this paper. In the case of \texttt{Small pebbles}, the melt mass fraction is smaller than that of Callisto with most of $\tau_{\rm form}$ and $t_{\rm start}$, because the impact velocity of the pebbles with ${\rm St}=0.01$ is reduced by the gas disk only at Ganymede. The critical Stokes number for Ganymede is ${\rm St_{crit}}=0.02$, larger than that of Callisto, because Ganymede is closer to Jupiter (Eq.(\ref{St-crit})).

\subsection{Ignored effects} \label{sec:ignore}
In this work, I only considered two heat sources, accretion heating and radiogenic heating by $^{26}$Al. However, there are many other heat sources such as radiogenic heating by other radioisotopes and hydration heat. These heat sources could cause more differentiation in Callisto after its formation even if it has not melted during the accretion.

Additionally, the effects of thermal blanketing are ignored in this work. \citet{joh23b} found that the interior of a protoplanet that grows by pebble accretion differentiates if the protoplanet grows large enough to hold on to a water vapor atmosphere, and the accretion luminosity is higher than the critical threshold for a run-away greenhouse effect, which is approximately $0.02$ Earth masses, just above the mass of Callisto. Therefore, if Callisto meets the conditions, the impact energy of pebbles may be maintained at the surface of the satellite and causes more differentiation than the results of this work.

These ignored heating effects may indeed help to explain the origin of the $\sim300~{\rm km}$ thick outer icy cluster of Callisto. However, the purpose of this research is to find necessary conditions of the accretion process of Callisto to maintain the partially differentiated interior. Therefore, such additional heat sources do not alter the conclusions of this paper; only pebble accretion has a chance to explain the partially differentiated interior of Callisto. The possibility of the convection of the interior can also be ignored for that purpose. This paper only focuses on the conditions for melting and does not matter what happens after the differentiation.

\section{Conclusions} \label{sec:conclusions}
``Planetesimal or pebble'' and ``satellitesimal or pebble'' have been discussed for long time, but clear answers have not been provided yet. In order to answer the fundamental questions, I focus on a unique large satellite of Jupiter, Callisto. Previous measurements of Callisto's gravity field suggests that the satellite is partially differentiated under the assumption of hydrostatic equilibrium \citep{and01}. I robustly showed that the interior is significantly differentiated by accretion heating with the satellitesimal accretion mechanism, since satellitesimals deposit some fractions of their impact energy under the surface of the satellite, which is consistent with the results of previous works \citep{bar08,ben25}. On the other hand, the interior can remain partially differentiated with the pebble accretion mechanism when the formation period is longer than $1.2~{\rm Myr}$, and the starting time of the accretion is late enough to avoid the melting by radiogenic heating. This is because 1) the impact energy of pebbles deposits at the surface of Callisto and effectively escapes by radiative cooling, and 2) the impact velocity of pebbles is reduced by the aerodynamic drag of the circumplanetary gas disk. Therefore, if the partial differentiation of Callisto is confirmed by future space missions such as JUICE \citep{cap22}, it could be the first observed evidence for the pebble accretion mechanism not only in the context of satellite formation but also in the broader framework of planet formation.

\begin{acknowledgments}
I appreciate the constructive comments by the anonymous referee, which improved this paper. I thank Masaki Takahashi and Shunichi Kamata for the very useful discussion about the long-term evolution of satellites' interiors. I also thank Yasuhito Sekine and Olivier Mousis for the exciting discussion about the idea of ``satellitesimal or pebble''. This work was supported by JSPS KAKENHI Grant Number JP24K22907.
\end{acknowledgments}


\appendix
\section{In the case of a lower initial ratio of aluminum-26 to aluminum-27} 
\label{sec:half-26Al}
I assumed the initial heating rate by $^{26}$Al as $q_{26}(0)=1.82\times10^{-7}~{\rm W~kg}^{-1}$, which is derived from the initial $^{26}$Al/$^{27}$Al ratio of $5.85\times10^{-5}$ \citep{bar08}. This value is appropriate for carbonaceous chondrites \citep{thr06}, and that is slightly higher than the canonical value, $5.23\times10^{-5}$ \citep{jac08}. However, some cosmochemical studies suggest that the outer Solar System had a lower initial value of $^{26}$Al/$^{27}$Al \citep[e.g.,][]{lar11,koo24}. Thus, I investigate the cases where the initial $^{26}$Al/$^{27}$Al ratio is half of the canonical value, corresponding to $q_{26}(0)=8.1\times10^{-8}~{\rm W~kg}^{-1}$.

\begin{figure*}[htbp]
\centering
\includegraphics[width=0.9\linewidth]{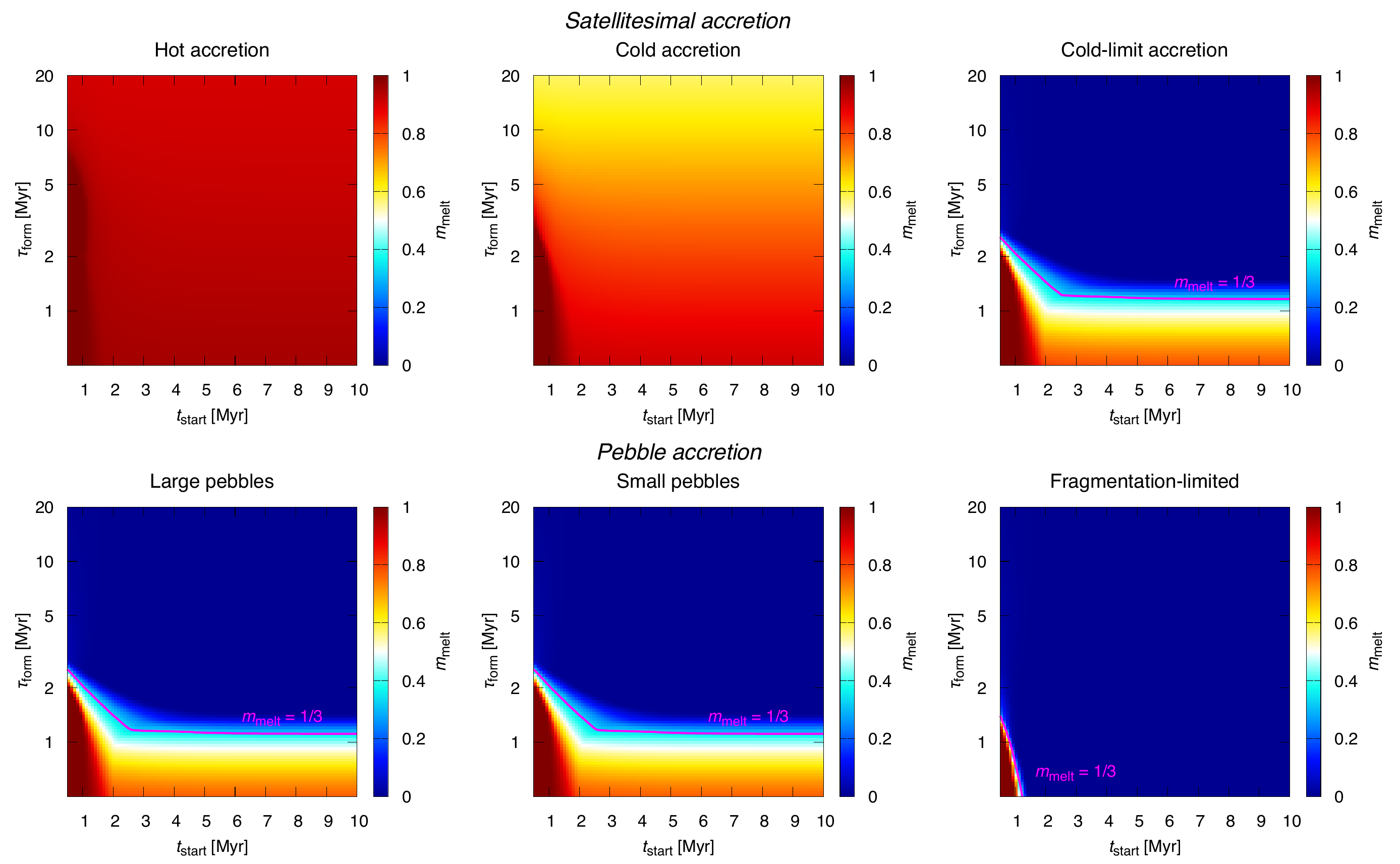}
\caption{Melt mass fractions of Callisto but for the lower initial $^{26}$Al/$^{27}$Al ratio.}
\label{fig:half-26Al}
\end{figure*}

Figure \ref{fig:half-26Al} shows that the effects of the lower initial $^{26}$Al/$^{27}$Al ratio is limited. The effective radiogenic heating regions (the left bottom corner of the panels) become smaller, but the other parts of the panels are the same as those of Fig. \ref{fig:meltmassfraction_Callisto}. Therefore, the conclusions of this paper do not change. The partially differentiated Callisto can only be achieved by pebble accretion even in the case of the lower initial $^{26}$Al/$^{27}$Al ratio.

\section{Melting point of ice-rock mixture in satellites} \label{sec:melting}
The interiors of satellites melt (differentiate) if the temperature becomes higher than the melting point of the material. Since the conditions of ice-rock differentiation is discussed in this work, the melting point of ice is necessary to be calculated. The melting point $T_{\rm melt}$ at $(R_{\rm int},t)$, depending on the pressure $P(R_{\rm int},t)$, is approximated by
\begin{equation}
T_{\rm melt}(P)=a+bP+c\ln{P}+d/P+e\sqrt{P},
\label{Tmelt}
\end{equation}
where $a$, $b$, $c$, $d$, and $e$ are coefficients for pure water ice listed in Table \ref{tab:Tmelt} \citep{dun10}. Effects of elements other than water are ignored. The pressure is calculated by
\begin{equation}
P(R_{\rm int},t)=\dfrac{2\pi}{3}G{\bar\rho}^{2}(R^{2}-R_{\rm int}^{2}).
\label{P}
\end{equation}

\begin{table*}[htbp]
\centering
\caption{Melting properties of water ice phases}
\hspace{-4.5em}
\begin{tabular}{lcccccc}
\hline
Ice phase & Pressure range [${\rm bar}$] & $a$ & $b$ & $c$ & $d$ & $e$ \\
\hline
Ice I & $P\leq2085.66$ & $273.0159$ & $-0.0132$ & $-0.1577$ & $0$ & $0.1516$ \\
Ice III & $2085.66<P\leq 3501$ & $10.277$ & $0.0265$ & $50.1624$ & $0.5868$ & $-4.3288$ \\
Ice V & $3501<P\leq6324$ & $5.0321$ & $-0.0004$ & $30.9482$ & $1.0018$ & $0$ \\
Ice VI & $6324<P\leq22160$ & $4.2804$ & $-0.0013$ & $21.8756$ & $1.0018$ & $1.0785$ \\
Ice VII & $P>22160$ & $-1355.42$ & $0.0018$ & $167.0609$ & $-0.6633$ & $0$ \\
\hline
\end{tabular}
\label{tab:Tmelt}
\end{table*}



\bibliography{Pebble-Callisto}{}
\bibliographystyle{aasjournalv7}

\end{document}